\documentclass[aps,nofootinbib,amsmath,prd,onecolumn,showpacs,superscriptaddress,groupedaddress,notitlepage]{revtex4-1}

\usepackage[american]{babel}
\usepackage{amsfonts}
\usepackage{amsmath}
\usepackage{amssymb}
\usepackage{epsfig}
\usepackage{latexsym}
\usepackage{paralist}
\usepackage{fancyhdr}

\usepackage{graphicx}

\usepackage[vcentermath]{youngtab}
\usepackage{young}
\usepackage{ytableau}
\usepackage{etex}
\usepackage{braket}
\usepackage{float}
\usepackage{slashed}
\usepackage{xcolor}
\usepackage{soul}
\usepackage{dcolumn} 
\usepackage{physics}
\usepackage{bm}
\usepackage{nicefrac}

\newcommand{\Diff}{{\rm \emph{Diff}}}
\newcommand{\Diffstar}{{\rm \emph{Diff}}^*}
\newcommand{\Weyl}{{\rm \emph{Weyl}}}
\newcommand{\Conf}{{\rm \emph{Conf}}}

\def\bea{\begin{eqnarray}} 
\def\eea{\end{eqnarray}}
\def\be{\begin{equation}} 
\def\ee{\end{equation}} 
\def\ba{\begin{array}}
\def\ea{\end{array}}

\def\be{\begin{equation}}
\def\ee{\end{equation}}
\def\bea{\begin{eqnarray}}
\def\eea{\end{eqnarray}}

\usepackage{amsmath}

\begin{document}

\title{The search for the universality class of metric quantum gravity}

\author{Riccardo Martini}
\email{riccardo.martini@oist.jp}
\affiliation{
Okinawa Institute of Science and Technology Graduate University, 1919-1, Tancha, Onna,
Kunigami District, Okinawa 904-0495, Japan}

\author{Alessandro Ugolotti}
\email{alessandro.ugolotti@uni-jena.de}
\affiliation{
Theoretisch-Physikalisches Institut, Friedrich-Schiller-Universit\"{a}t Jena,
Max-Wien-Platz 1, 07743 Jena, Germany}

\author{Omar Zanusso}
\email{omar.zanusso@unipi.it}
\affiliation{
Universit\`a di Pisa and INFN - Sezione di Pisa, Largo Bruno Pontecorvo 3, I-56127 Pisa, Italy}

\begin{abstract}
%
On the basis of a limited number of reasonable axioms, we discuss the classification of all the possible
universality classes of diffeomorphisms invariant metric theories of quantum gravity.
We use the language of the renormalization group and adopt several ideas
which originate in the context of statistical mechanics and quantum field theory.
Our discussion leads to several ideas that could affect the status of the asymptotic safety conjecture
of quantum gravity and give universal arguments towards its proof.
\end{abstract}

\maketitle

\renewcommand{\thefootnote}{\arabic{footnote}}
\setcounter{footnote}{0}

\section{Introduction}\label{sect:introduction}

A well-known fact is that in four dimensions a quantum theory of Einstein-Hilbert gravity
displays perturbatively non-renormalizable divergences starting at two loops \cite{Goroff:1985th,vandeVen:1991gw}.
This happens because Newton's constant has negative mass dimension, which in turn requires a re-interpretation
of the perturbative series as an effective expansion in inverse powers of the Planck mass, making
the theory an effective IR model, rather than a UV complete one \cite{Donoghue:1994dn}.

It was suggested by Weinberg that the UV limit of metric gravity
could still be meaningful if the theory was asymptotically safe, namely, if it had
an ultraviolet fixed point with a finite number of relevant directions in its renormalization group flow \cite{Weinberg:1976xy}.
Since its inception, the idea of asymptotically safe gravity was linked to the proposal of
continuing two-dimensional gravity, which is asymptotically free, in $d=2+\epsilon$ dimensions \cite{Kawai:1989yh},
where it could be asymptotically safe \cite{Weinberg:1980gg}.

The proposal was early on investigated perturbatively continuing from two dimensions \cite{Jack:1990ey,Aida:1996zn}, but,
given the obvious limitations of the $\epsilon$-expansion,
the main tool to explore the conjecture soon became the non-perturbative functional renormalization group,
starting from the pivotal work of Reuter \cite{Reuter:1996cp}.
One difficulty, which is caused by the use of the average effective action of the functional approach,
is that it is sometimes difficult to eliminate parametric and gauge dependences from the action, see for example \cite{Gies:2015tca},
making the physical interpretation of the results unclear.
To this day, the possibility that quantum gravity is asymptotically safe is still a conjecture,
and most theoretists working on quantum gravity have strong opinions on the matter, either in favor or against it \cite{Donoghue:2019clr,Bonanno:2020bil}.

The discussion on the early approach to the conjecture brings forward a dilemma:
on the one hand, using perturbation theory, we have that gravity in $d=2+\epsilon$ is a well-defined
theory with no gauge or parametric dependence on-shell, but we are limited to $\epsilon \ll 1$, so we cannot access
the physically interesting four dimensional theory located at $\epsilon=2$ (a meaningful resummation would require many more orders
than the available ones). On the other hand, the non-perturbative approach has access to $d=4$ directly,
but has general scheme and background/gauge dependences and no obvious way to eliminate them.
The natural question is, which method should one choose?
Our suggestion is simple, both. The search for the ultraviolet completion of metric gravity
should be pursued on both the perturbative and non-perturbative sides, and both sides
should learn lessons from each other.

If we are to accept the idea that the UV complete theory of quantum gravity in $d=4$ comes from
the model in $d=2+\epsilon$, which is an idea based entirely on perturbation theory, then we might ask
whether there are obstructions to the continuation of this solution to $\epsilon=2$.
In this paper, we discuss possible mechanisms that could either allow or prevent the $\epsilon$-expansion
from reaching the physical case $d=4$. One natural mechanism is that of collision of fixed points,
which is difficult to capture perturbatively, unless all involved players are in the weak-coupling regime.
For this reason, it is important to have an idea of all the possible critical theories that are described by a
metric effective action.

In the next two sections we use general arguments to discuss metric-based diffeomorphisms invariant
theories that can be constructed perturbatively at a given critical dimension $d_{cr}$.
We also argue if and how these theories can be continued either below or above their critical dimension
to the physically interesting case $d=4$.

\section{The diffeomorphisms group and its siblings}\label{sect:diffeomorphisms}

The first ingredient of our construction requires the choice of the symmetry group.
There are in principle several possible options of symmetry groups that lead to sensible gravitational actions.
A discussion on some possibilities and the degrees of freedom that they propagate can be found in \cite{Gielen:2018pvk}.
In order to contain the proliferation of possibilities, we mostly limit ourselves to the diffeomorphisms group and close relatives.

Einstein's general relativity is an example of theory which is invariant under general diffeomorphisms.
Infinitesimally, the general transformation can be parametrized locally with a vector field $\xi^\mu$,
so that coordinates change nonlinearly as $x^\mu \to x^{\prime\mu}=x^\mu+\xi^\mu(x)$.
The metric transforms as
\begin{equation}\label{eq:diff-action}
\begin{split}
 \delta_\xi g_{\mu\nu} &= \nabla_\mu \xi_\nu+ \nabla_\nu \xi_\mu\,.
\end{split}
\end{equation}
These infinitesimal transformations generate the algebra of the diffeomorphisms group, denoted $\Diff$,
which is isomorphic to the one of vector fields with Lie brackets as internal commutator
\begin{equation}\label{eq:diff-algebra}
\begin{split}
 \left[\delta_{\xi_1},\delta_{\xi_2}\right] &= \delta_{\left[\xi_1,\xi_2\right]}\,,
\end{split}
\end{equation}
where $\left[\xi_1,\xi_2\right]^\mu=\xi_1^\nu \partial_\nu \xi_2^\mu -\xi_2^\nu \partial_\nu \xi_1^\mu $
are the Lie brackets of two vectors.

If one chooses to rewrite the metric as a conformal factor times another metric, $g_{\mu\nu}=\varphi^{\nicefrac{4}{d-2}} \tilde{g}_{\mu\nu}$,
the symmetry is generally enhanced by a Weyl factor related to the combined rescaling of the two terms. By requiring that
transformations on $\tilde{g}_{\mu\nu}$ preserve the volume form $\sqrt{\tilde{g}}$,
it is possible to break the enhanced symmetry down to a subgroup, $\Diffstar$, for which
\begin{equation}\label{eq:diff-unimodular-dilaton-action}
\begin{split}
 \delta^*_\xi \tilde{g}_{\mu\nu} = \tilde{\nabla}_\mu \xi_\nu+ \tilde{\nabla}_\nu \xi_\mu -\frac{2}{d}\tilde{g}_{\mu\nu} \tilde{\nabla}_\alpha \xi^\alpha
 \,,
 &\qquad\qquad
 \delta^*_\xi \varphi = \xi^\mu  \partial_\mu \varphi +\frac{d-2}{2d} \varphi\, \tilde{\nabla}_\alpha \xi^\alpha
 \,,
\end{split}
\end{equation}
where indices are lowered with $\tilde{g}_{\mu\nu}$ on the right hand side. The subgroup preserves the volume, therefore $\tilde{g}_{\mu\nu}$
is a unimodular metric, while $\varphi$ is referred to as the dilaton. The interesting part is that this symmetry group is isomorphic to $\Diff$ itself,
$\Diff\simeq\Diffstar$ as seen from proving that
\begin{equation}\label{eq:diffstar-algebra}
\begin{split}
 \left[\delta^*_{\xi_1},\delta^*_{\xi_2}\right] &= \delta^*_{\left[\xi_1,\xi_2\right]}\,.
\end{split}
\end{equation}
A $\Diffstar$ transformation can be seen as a volume preserving diffeomorphism combined to a conformal transformation, which together
return a general diffeomorphism on the original metric. They propagate the same degrees of freedom, but the two groups generally differ when
considered in a path-integral.

Another possiblity would be to not break the Weyl factor and construct a theory that is invariant under
a generalization of the conformal group 
$\Conf = \Diff \ltimes {\Weyl}$ and would require the theory to be invariant under both Eq.~\eqref{eq:diff-action}
and the local rescalings of the form $g_{\mu\nu} \to \Omega^2 g_{\mu\nu}$. By construction, theories invariant under $\Conf$ have conformally invariant solutions
of the equations of motion, so are themselves invariant under the conformal group, which is why,
with an abuse of notation we refer to $\Diff \ltimes {\Weyl}$ as $\Conf$.
We refer to \cite{Karananas:2015ioa} for an overview of the difference between $\Weyl$ and $\Conf$.
In a conformally invariant theory one does not naturally have scales, such as masses, therefore,
in order to reproduce the physical world that includes Newton's constant for gravity,
one should invoke a spontaneous breaking mechanism at lower energies \cite{tHooft:2011aa}.
One point has to be made on the conformal group, and in particular on the $\Weyl$ subgroup,
as it does not seem to have a nonzero Noether current in simple scalar models,
which could make one question
its role in the construction of a dynamical theory \cite{Jackiw:2014koa}. This is a particularly delicate
issue that deserves more attention.

The symmetries that we described here are but two of a more general list of examples discussed in Ref.~\cite{Gielen:2018pvk}.
In fact, specific breaking patterns of the conformal group actually generate the aforementioned $\Diff$ and $\Diffstar$ \cite{Gielen:2018pvk,Martini:2021slj},
so we could reasonably expect them as outcomes of obtaining Planck mass at low energies from conformal symmetry.
Another natural symmetry that emerges in this context is the group of volume preserving diffeomorphisms, that is the symmetry of unimodular gravity which has been introduced to relax the tension between the values of cosmological and Newton's constants (see for example Refs.~\cite{Benedetti:2015zsw,deBrito:2020rwu}).

\section{Guiding principles and illustrative examples}\label{sect:construction}

Now we articulate the guiding principles, inspired by the application of the renormaliation group
to the theory of critical and multicritical phenomena in a Ginzburg-Landau description,
that we are then going to apply to the case of metric-based theories.
Some of the following ideas might be familiar to all readers, in that they are based on the textbook
examples of $\phi^4$ and Yang-Mills gauge theories. Some others might be less familiar,
especially those based on less-traditional examples, so we hope that they are met by open-minded readers.
Rather than appliying directly any idea to the metric case, we prefer to discuss the principles using
as many examples coming from the field-theoretical approach to statistical mechanics as possible.

\bigskip

\noindent \fbox{
\parbox{0.972\textwidth}{
{\bf Guiding principle 1:} The theory admits a \emph{critical dimension} $d_c$ at which a certain finite set of
operators ${O_i}$ are canonically marginal.
}
}

\bigskip

\noindent Notice that we do not assume here that $d_c$ equals four, even though it may do so.
The idea behind this principle is that from the set of operators $O_i$ we can
construct an action $S= S_{\rm free} + \int{\rm d}^{d_c} x \sum_i g_i O_i$, which is parametrized by the set of couplings $g_i$ that,
by definition, have zero mass dimension $[g_i]=0$. The first term, $S_{\rm free}$, is some opportune
free action that has no coupling, which can be used to fix the canonical dimensions of the fields,
and consequently of the couplings $g_i$.
The operators $O_i$ could be all possible operators compatible with a given symmetry for the model.
On this couplings' basis, the theory is power counting renormalizable at $d=d_{c}$, that,
to state the obvious is \emph{not necessarily} equal to four.
We also do not assume that $d_{c}$ is the upper critical dimension of the theory, which instead we denote $d_{u}$. The upper critical dimension $d_u$ 
is defined as the dimension above which the theory is Gaussian. It might as well be $d_c \leq d_u$, as we discuss below.\footnote{%
There can even be a lower critical dimension, but we assume that it is sufficiently low in all the examples below.
}

\bigskip

\noindent \fbox{
\parbox{0.972\textwidth}{
{\bf Guiding principle 2:} The operators $O_i$ govern a well-defined renormalizable perturbative expansion at the critical dimension $d=d_c$,
which expresses everything as an expansion in powers of the couplings $g_i$.
}
}

\bigskip

\noindent This is the essential requirement of perturbative renormalizability of $S$ at the critical dimension.
Since the theory is perturbatively renormalizable all divergences are local and have the structure of the operators $O_i$.
Divergences can be eliminated
through the introduction of the finite number of counterterms order-by-order
in the perturbative expansion.
Generically, we expect that, if we use the method of dimensional regularization
and the modified minimal subtraction scheme $\overline{\rm MS}$, all counterterms will
result in renormalized couplings $g_i$, denoted with the same symbol as the bare ones for simplicity,
which make the path integral finite. If $\mu$ is the scale of dimensional
regularization at which the divergences are subtracted, then the couplings will flow
with beta functions $\mu \partial_\mu g_i(\mu)= \beta_i(g)$.
With only one exception given in Sect.~\ref{sect:YM}, all the beta functions discussed in this paper
should be understood as being derived using minimal subtraction
and dimensional regularization methods.
We also reserve the symbol $\beta$ for the running of the couplings at $d=d_c$ (see also footnote \ref{footnote:betas}).

\bigskip

\noindent \fbox{
\parbox{0.972\textwidth}{
{\bf Guiding principle 3:} The renormalized theory can be continued to a \emph{finite} interval of $d$ which includes $d_c$.
}
}

\bigskip

\noindent The continuation of the renormalized theory in $d=d_c$ to another dimension could begin with an
$\epsilon$-expansion in $d=d_c-\epsilon$, however the emphasis on the latter principle is on the word ``finite'', 
so it is assumed that the theory can be meaningfully extended to finite values of $\epsilon$. This finite value, say $d_u$, can
still be very small, though not infinitesimal; in fact, it might not even intercept any physically interesting value for $d$,
which is a possibility that we have to deal with, in some way.

\bigskip

Even though we are interested in finite extensions away from $d_c$, 
still the $\epsilon$-expansion gives us particularly strong insights on how and where the theory could be extended in $d$.
To elaborate on this point, consider the action $S$, which is traditionally extended to $d=d_c-\epsilon$ dimensions as
$S= S_{\rm free}+\int{\rm d}^{d} x \, \sum_i \mu^{a_i\epsilon} g_i O_i$, where we include the scale $\mu$ and some positive contants $a_i$
in such a way that the couplings remain dimensionless.
For infinitesimally small $\epsilon$, this implies that the couplings now obey the new scale transformations,
$\mu \partial_\mu g_i(\mu) = -a_i\epsilon g_i + \beta_i(g)$ (no summation over $i$), that now include a scaling term.\footnote{%
Here and in the following, we reserve the symbol $\beta_i(g)$ for the beta function of the coupling $g_i$ at the critical dimension,
while the general beta function in $d$ dimensions is denoted $\mu \partial_\mu g_i(\mu)$.
For example, a scalar $\phi^4$-theory with coupling $\lambda$ has $d_c=4$
and requires the subtraction from the bare coupling of a counterterm of the form
$\frac{A}{\epsilon}\lambda^2 \mu^{-\epsilon}$ for some positive constant $A$
when the theory is regularized in $d=4-\epsilon$ dimensions \cite{Kleinert:2001hn}. In the loop expansion, $A$
is determined from three one-loop diagrams with four external legs pairwise inserted in the loops.
The scale $\mu$ appears as a byproduct of continuing the theory away from $d=4$, where the coupling is
no longer dimensionless, but, instead, has dimension $\mu^\epsilon$
($a_i=1$ in the notation of the main text).
The independence of the bare coupling from the scale $\mu$ implies that the renormalized coupling
runs as $\mu\partial_\mu \lambda = A \lambda^2$ when $d=4$
(both sides are correctly dimensionless at the upper critical dimension),
which defines $\beta_\lambda = A \lambda^2$.
Instead, in $d=4-\epsilon$, it is customary to measure $\lambda$ in units of $\mu$ in order
to still have a dimensionless coupling.
This is achieved through the replacement $\lambda \to \mu^\epsilon \lambda$,
as shown in the main text. The counterterm in the new units becomes $\frac{A}{\epsilon}\lambda^2 \mu^{\epsilon}$
and the requirement that the bare coupling is independent of $\mu$ translates to
a running of the renormalized one,
$\mu\partial_\mu \lambda = -\epsilon \lambda +A \lambda^2= -\epsilon\lambda +\beta_\lambda$,
which correctly interpolates the limit $d=d_c$ for $\epsilon\to 0$.
\label{footnote:betas}
}
To better understand the implication of the new running, we momentarily restrict our attention
to a single coupling $g$, which has a polynomial beta function $\beta$. We also assume that the leading term of the beta function
in $d=d_c$ is $\beta \sim g^2$ and that $g>0$ is required for boundedness,
although these requirement can be dropped with minor modifications on the final arguments.
The important point is that, according to the sign of $\beta$, the model can be either asymptotically free ($\beta<0$) or Landau-trivial ($\beta>0$).
This happens because for $\epsilon \neq 0$, according to the signs of $\beta$ and $\epsilon$,
there can be an interesting scale invariant fixed point $g^*$ defined implicitly as a solution of $a \epsilon g^* = \beta|_{g=g^*}$.
\begin{itemize}
 \item If $\beta>0$, that is, if the theory is Landau-trivial at $d=d_c$,
 we have that $g^* \sim {\mathcal O(\epsilon)}$ is a physically interesting fixed point for $\epsilon>0$,
 implying that the theory could be extended nontrivially to $d<d_c$.
 In this case, $d_c$ coincides with the \emph{upper} critical dimension, $d_c=d_u$.
 The natural interpretation is that the fixed point $g^*$ is infrared, because it governs the scale dependence of the model at low energies, $\mu\to 0$,
 that can be seen from a simple stability analysis. When $d=d_c$ the model is trivial, meaning that the IR is governed by the Gaussian point, with at
 most logarithmic corrections to scaling. An example of Landau-trivial theory would be $\phi^4$ in $d_c=4$;
 \item If $\beta<0$, that is, if the theory is asymptotically free,
 we have that $g^*$ is physical for $-\epsilon>0$. This suggests that the theory can be extended nontrivially to $d>d_c$,
 mirroring the previous case, at least from a formal point of view.
 In this case, we can assume that $d_u > d_c$, however we do not know if $d_u$ is big enough to include interesting physical values.
 The mirroring continues in that the fixed point $g^*$ is ultraviolet, because it governs the scale dependence of the model at high energies, $\mu\sim \infty$.
 An example of asymptotically free theory would be a $SU(N)$ gauge theory in $d_c=4$.
\end{itemize}

\begin{figure}[tb]
\includegraphics[width=0.45\textwidth]{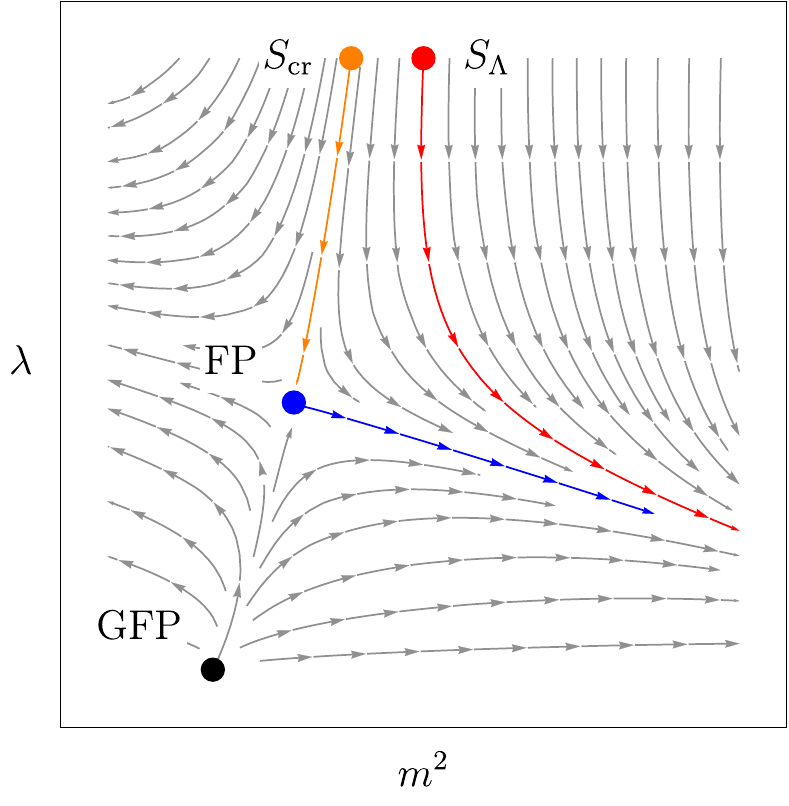}
\caption{RG diagram illustrating the relation between the linear and nonlinear sigma models; the trajectories represent lines of constant physics at large scales.
The couplings are a quartic self interaction of the linear model and a squared mass in units of the RG scale,
and their RG system has a nontrivial fixed point (FP) that is distinguished from the Gaussian fixed point (GFP).
The fixed point controls the \emph{infrared} behavior of the linear sigma model, as can be seen from the red trajectory that spawns from the point $S_\Lambda$
that represents a given ``bare'' theory characterized by a microscopic scale $a$ or an ultraviolet energy $\Lambda\sim a^{-1}$.
The fixed point drives the red trajectory to some
infrared limit, located on the right in this example, that is representative of a macroscopic phase (e.g.\ broken or unbroken symmetry).
In statistical mechanical models, $S_\Lambda$ can be tuned to a critial value $S_{\rm cr}$ (for example moving it towards the left in the picture), so that the
trajectory would fall towards the fixed point itself following the orange trajectory
and the theory becomes scale invariant by construction.
The tuning is often achieved by changing the temperature to its critical value, $T\to T_c$,
in the thermodynamical limit and under the assumption that the parameters of $S_\Lambda$ depend on $T$.
The same fixed point, however, is seen as an \emph{ultraviolet} one for the theories located on the blue trajectory,
which represent the nonlinear sigma model. In the infrared limit the red and blue trajectories are taken to the same
macroscopic phases, because they follow the same RG directions asymptotically, and thus belong to the same universality class.
One important message is that the two models, linear vs nonlinear, see the same fixed points in two rather different lights,
infrared vs ultraviolet, however it is perfectly legitimate to talk about the notion of universality class for both models.
A useful remark for gravity is the following: an asymptotically safe trajectory for gravity would be the blue curve, since it is ultraviolet complete,
while the rest of the RG diagram could represent any lattice-based attempt to find the continuum limit of gravity
based on some lattice scale $a\sim\Lambda^{-1}$ (e.g.\ any approach based on Regge's discretization \cite{Regge:1961px,Hamber:2007fk},
such as EDT or CDT \cite{Ambjorn:2000dv,Coumbe:2014nea}),
in which case the ultraviolet critical point for quantum gravity is seen as a standard fixed point associated to a second order phase transition in the lattice
(see also the appropriate discussion on Sect.~\ref{sect:YM}).
\label{fig:nlsmuniversality}}
\end{figure}

A well-known example of a theory for which $\beta>0$ and that can be continued
below its critical dimension is the Wilson-Fisher $O(N)$ model with $\phi^4$ interaction, also known as the linear sigma model (LSM) \cite{Wilson:1971dc}.
In fact, the $\phi^4$ interaction is Landau-trivial at $d_c=4$, but has a nontrivial fixed point for $2\leq d < 4$,
which encapsculates the large-scale behavior
of a universality class of models which includes the three-dimensional lattice Ising model for $N=1$.
An example of theory for which $\beta <0$ and that can be continued above its critical dimension is the $O(N)$ nonlinear sigma model (NLSM) \cite{Brezin:1975sq}.
The NLSM is asymptotically free in $d_c=2$ and can be continued to $d>2$ to describe a critical point which is believed to be the same as
the LSM universality class for $N>2$. One intuitive way to understand why they should belong to the same universality class is to realize that the NLSM
can be interpreted as a LSM in which the radial mode $\rho\sim\sum_i\phi_i^2$ has been integrated-out, or, alternatively, the LSM can be interpreted
as a NLSM in which the radial mode reappears as a bound state \cite{Bardeen:1976zh}. The relation between LSM and NLSM can be visualized in Fig.~\ref{fig:nlsmuniversality}.
Since the two models share the same universality, it is implied that they should share the upper critical dimension,
which in the case of the Wilson-Fisher is $d_u=4$. There is thus a finite interval of existence for the critical point of the NLSM above $d_c=2$,
even though a naive extrapolation of the leading order of the $\epsilon$-expansion would suggest otherwise.

In fact, the examples of LSM versus NLSM illustrate a number of interesting points. The first one is that different models may share
the same universality class at a critical point, which is a well-known fact of the theory of critical phenomena.
Another one is that, from the point of view of the renormalization group,
infrared and ultraviolet are relative concepts. Loosely speaking, one could imagine the critical point of the LSM in $2\leq d < 4$ as lying
on an RG phase diagram parametrized by mass and self-interaction. The RG flow would then swing close to the fixed point,
because there are a relevant and an irrelevant directions, approximately identified with mass and self-interaction, respectively. The swung trajectories
are then collected by the RG and driven to some infrared phase with macroscopic interpretation
(for example zero or non-zero magnetization in the case of the sigma models),
giving to the fixed point a natural infrared interpretation. However, in the same plane it is possible to identify the unique trajectory
that departs from the critical point towards the infrared, that is, the infrared-relevant direction. In this case, the fixed point has a natural ultraviolet interpretation
in agreement with the point of view of the analysis of the NLSM.

This picture that we just described is shared qualitatively by other models \cite{Janssen:2016xvc}.
Another interesting and rather general aspect is that for models with a sufficiently large number of scalar degrees of freedom,
but presumably for many other models too, it can actually be shown that there is no absolute infrared
fixed point in the most general RG space of theories \cite{Osborn:2020cnf}.

All the above considerations can be generalized to the case of multiple $g_i$ couplings
if the opportune care is given.
In general, the action $S$ will have to satisfy some positivity conditions on the couplings in order to generate a meaningful
path-integral. Furthermore some additional constraints could be imposed on the basis of unitarity for Lorentzian models,
reflection positivity for Euclidean models \cite{Arici:2017whq}, and stability.
On the basis of accumulated experience, we expect that a general solution,
which comes from the set of fixed point equations,
can be either continued above or below $d_c$, or maybe even both.
For a meaningful theory of gravity the crucial question is, for obvious reasons, whether it can be continued to $d=4$.

The $\epsilon$-expansion, if combined with the sign of the perturbative beta function at $d_c$, guarantees that a fixed point can be continued at least to some infinitesimally small
value $d=d_c-\epsilon$ and $\left|\epsilon\right|\ll 1$. From our point of view this in enough to begin speculating on the continuation of perturbative fixed points as reasonable
candidates of quantum theories of gravity. We also want to conjecture on which could be the actual mechanisms that prevent a theory from being useful for finite values of $\epsilon$.
In the assumption that fixed points do not simply ``come and go'', but that there must be a
mechanism that makes them appear and disappear, we are led to including another guiding principle.

\bigskip

\noindent \fbox{
\parbox{0.972\textwidth}{
{\bf Guiding principle 4:} Distinct critical points do not simply exist in a vacuum: they can collide with each other, resulting, for example, in complex conjugate pairs as a function of the parameters of the theory.
}
}

\bigskip

\noindent This is by far our most speculative statement, so we are urged to try and justify it. There is a well-known statistical model, the lattice $q$-states Potts model,
which is believed to display an annihilation of fixed points as a function of the number of states $q$ \cite{Gorbenko:2018dtm}. In the continuum, the universality class of the Potts model is described by
the Landau-Potts model, which is a scalar model with $\phi^3$ interaction constructed to be invariant under $S_q$ symmetry.
The Landau-Potts theory admits a perturbative expansion in $d_c=6$,
and an $\epsilon$-expansion in $d_c=6-\epsilon$. Likewise the $\phi^4$ theory,
the critical point exists and can be continued up at least to $d=2$.
In the range $2\leq d \leq 6$, it is assumed to satisfy a condition in the number of states,
that is, there must be a critical value of the number of states, $q_c$,
which separates a second order phase transition for $q<q_c$
from a first order phase transition for $q>q_c$.

For example, in $d=2$ it is well-known that the critical value of $q$ is $q_c=4$ \cite{Gorbenko:2018dtm}.
It has been conjectured that in arbitrary $d$, the critical value of the number of states
must be $d$ dependent, thus drawing a curve $q_c=q_c(d)$ in the $(q,d)$ plane \cite{Codello:2020mnt}.
The different order of the phase transition stems from the presence or absence of
the nontrivial fixed point as a function of $q$: if there is a nontrivial \emph{real} fixed point
the standard argument of phase transitions is that there is a second order phase, if instead
there is only the Gaussian fixed point the system undergoes a first order phase transition.

Perhaps more interesting is the mechanism with which the theory's transition changes from second to first order as a function of $q$, which is well-established in $d=2$ and conjectured to hold in any dimension.
There are actually two fixed points, a critical and a multicritical one,
that are distinct for $q<q_c$, but that collide at $q=q_c$ and then become
a complex conjugate pair for $q>q_c$, leaving only the Gaussian fixed point.
This mechanism has the added value of explaining why, for some values of $q$
close to and above $q_c$, the transition is weakly first order: if the two fixed points
just collided the complex conjugate pair that is left makes the transition
almost second order because it is still ``close'' to the real axis, in some sense that must be made precise \cite{Gorbenko:2018dtm}.

Several analyses of perturbative renormalization groups see fixed points that collide for some parametric value. An example could be the analysis of the hypercubic model in $d=4-\epsilon$, which has the symmetry group of the hypercube $H_N$,
in which it is easy to see that a fixed point with genuine hypercubic symmetry and the isotropic fixed point with $O(N)$ symmetry collide at a value $N_c$. In the hypercubic example, the critical value $N_c$ can be determined as a function of $\epsilon$, hence of the dimension,
and, from careful analysis, is very close to $N_c\approx 3$ for $d=3$ (perturbatively, it is $N_c=4$ at $d=d_c=4$,
and it is a difficult to estimate that $N_c < 3$ in $d=3$ from perturbation theory) \cite{Pelissetto:2000ek}.
The hypercubic example is interesting, but somehow simplifies the general case
because both the $O(N)$ and $H_N$ critical points ``live'' in the same perturbative RG diagram, meaning that their perturbative series can be described by the same set of RG beta functions.

Returning to the example of the Potts model, we see that the annihilation of fixed points is even less simple in this case, precisely for the reason that we have just mentioned. While the Landau-Potts theory is governed by the $\phi^3$ interaction in $d=6-\epsilon$, the multicritical parter, with which the standard Potts annihilates, is not.
A reasonable conjecture is that, in a Landau free-energy description, such multicritical model
is described by a higher order interaction, with the simplest candidate pushed forward being $\phi^5$ in $d=\nicefrac{10}{3}-\epsilon$.
The point is that the critical and multicritical models have different $d_c$, $6$ and $\nicefrac{10}{3}$, respectively, so it is impossible to see them in the same perturbative RG diagram. In fact, they are governed by two different perturbative series, or, in other words,
they can at most appear in the strongly-interacting regime of each other \cite{Codello:2020mnt}.

We stress that there are several other interesting examples of colliding fixed points in gauge theories (e.g.\ see \cite{Kaplan:2009kr}, which we return to in Sect.~\ref{sect:YM} from an unusual perspective.
These mechanisms generally lead to \emph{conformal windows}
that depend on the number of flavors,
but we anticipate that these ideas resonate also
with part of the discussion that we are going to give in Sect.~\ref{sect:2d}, where the number of flavors is replaced by the dimension of the system.
Among these, we find particularly important the discussions
that lead to the dynamical symmetry breaking
in three-dimensional QED \cite{Kubota:2001kk,Kaveh:2004qa}
and the chiral symmetry breaking of QCD with many flavors \cite{Gies:2005as}.

\section{The candidate metric theories}\label{sect:candidates}

For the construction of the possible candidate metric theories that satisfy the first guiding principle,
we assume that the only field at our disposal is the metric $g_{\mu\nu}$, that the connection is the Levi-Civita one,
and that the theory must be covariant either under $\Diff$ or $\Diffstar$. We also choose to work with
covariant tensors that are constructed from the metric, namely the Riemannian curvatures,
and that can be defined unambiguously in any dimension, for obvious reasons.

If we engineer coordinates to be dimensionless, then the metric tensor has mass dimension negative two, canonically, like a length square.
The Riemann tensor with three indices down and one up, $R_{\mu\nu}{}^\alpha{}_\beta$,
is dimensionless; likewise the Ricci tensor is dimensionless $R_{\mu\nu}=R_{\alpha\mu}{}^\alpha{}_\nu$;
finally the curvature scalar has mass dimension two being contracted with an inverse metric, $R=g^{\mu\nu}R_{\mu\nu}$.
These three tensors, together with the covariant derivative $\nabla_\mu$, are the covariant quantities that we have at our disposal to construct an action.

It is easy to see that operators constructed with $n$ powers of the curvatures, such as for example $R^{n}$ have dimension $2n$, while all lower powers of the curvatures could be studied as composite operators \cite{Houthoff:2020zqy}.
If we take into account the invariant volume element, $\sqrt{g}$, then the couplings multiplying an $n$-th power of the curvatures
have canonical dimension $d-2n$. To satisfy the first principle, we seek for canonically dimensionless couplings at $d_c$,
which imply the solutions $d_c=2n$. The application of the first principle thus leaves us with the even natural numbers
as candidate theories to look forward to. In the next sections we briefly review the work that has been done for the first few examples
and collect some additional ideas.

\subsection{Gravity in $d=2+\epsilon$}\label{sect:2d}

We have that the only available local operator built from the metric of dimension two is the curvature scalar,
so in two dimensions the gravitational coupling itself, the Newton's constant, becomes dimensionless and
must be used to construct a perturbative expansion. In the simplest units, the action is
\begin{equation}\label{eq:action2d}
\begin{split}
 S[g] &= - \frac{1}{G}\int {\rm d}^2 x \sqrt{g} R\,,
\end{split}
\end{equation}
and,
close to two dimensions, it is easy to see that the perturbative coupling is $G$ by expanding the
metric as
$\bar{g}_{\mu\nu}=\delta_{\mu\nu}+G^{\alpha}h_{\mu\nu}$ for some $d$-dependent power $\alpha$
and a canonically normalized symmetric tensor fluctuation $h_{\mu\nu}$.
However, in two dimensions several things happen at the same time: the curvature becomes a topological invariant when integrated, the action
is conformally invariant modulo boundary terms, and the Lagrangian is zero on-shell.
The handling of these problems results in a rather different way of treating the conformal mode between the $\Diff$ and $\Diffstar$ realizations.
In the first case, the action \eqref{eq:action2d} is complemented with a cosmological constant term that is renormalized simultaneously and allows to go on-shell
consistently. In the second case, the conformal mode is separated and the corresponding scalar-tensor theory is renormalized. The $\Conf$ invariant result can be evinced easily from String Theory in that it is related to the leading contribution to the string's dilaton effective action in a non-critical dimension.

Either way, at the leading order the beta function for Newton's constant is of the form
\begin{equation}\label{eq:beta2d}
\begin{split}
 \beta_G&=\frac{c_g}{24\pi}G^2\,,
\end{split}
\end{equation}
and $c_g=-19$ for $\Diff$ \cite{Jack:1990ey}, $c_g=-25$ for $\Diffstar$ \cite{Polyakov:1987zb},
and $c_g=-26$ for $\Conf$ \cite{Polyakov:1981rd,Odintsov:1991qu,Elizalde:1995gw},
which can be obtained from the coupling of \eqref{eq:action2d}
to an additional scalar. The Newton's constant is obviously asymptotically free in $d=2$
for the pure gravity model, while there is a conformal window if matter fields are included in the problem.
In $d=2+\epsilon$, following the discussion of the previous section, the theory is asymptotically safe because it has a fixed point, $G^*\sim {\cal O}(\epsilon)$,
however we do not know how far this fixed point can actually extend.

The combination of functional renormalization group (FRG) and background approach, with some approximations,
seems to suggest that such fixed point exists up to $d=\infty$ \cite{Litim:2003vp}.
This fact, however, is not backed by a computation that discusses all possible parametrization dependences in the construction \cite{Gies:2015tca},
so we argue that it is an artifact of the approximation made in background projection
(in fact, the same effect can be seen in the application of FRG to the NLSM \cite{Codello:2008qq}, but we know that
the NLSM universality class has upper critical dimension equal to four).
In a previous work, we have discussed that there should be a conformal window in $d$, which could be thought of
as a parameter analog to $N$ in $SU(N)$ gauge theories, and we have also given a reliable estimate for the size of the window
in the case of $\Diff$ symmetry that agrees with previous cutoff-based work.\footnote{%
It would be interesting, and certainly fruitful for the entire discussion of this paper, to formalize the continuation to arbitrary continuous dimension $d$.
A possible starting point, at least for the Euclidean theory, could be the method discussed in Ref.~\cite{Binder:2019zqc},
in which it is made sense of $O(n)$ ``symmetry'' for arbitrary values of $n$,
however it is still an open problem to generalize that method to gauge symmetries.
}
The $d$-dependent beta function becomes
\begin{equation}\label{eq:betaG-JJ}
\begin{split}
 \beta_G&=\epsilon G - \frac{36+3d-d^2}{48 \pi } G^2\,,
\end{split}
\end{equation}
and we find that the fixed point exists for $d\lesssim 7.5$, modulo further corrections from higher loops,
which seems reasonably bigger than four to justify the conjecture of asymptotic safety in $d=4$ \cite{Falls:2015qga}.

It is natural to ask which is the mechanism that would make this fixed point disappear. In the NSLM example the fixed point
collides with the Gaussian one at $d=4$, so the same could happen to the asymptotically safe fixed point of $d=2+\epsilon$ gravity.
Another possibility, that we would like to put forward here, is that this fixed point collides and annihilates with the one of a higher derivative
generalization of \eqref{eq:action2d}. This last possibility is of course purely conjectural, but it is the reason why we spend some words
on the next two higher derivative models of gravity.

\subsection{Higher derivative gravity in $d=4$}\label{sect:4d}

In four dimensions the couplings multiplying operators that are quadratic in the curvatures are dimensionless,
and thus suitable to construct a perturbatively renormalizable quantum theory. The action is often parametrized as
\begin{equation}\label{eq:action4d}
\begin{split}
 S[g]&=
 \int {\rm d}^4 x \sqrt{g} \Bigl\{
  \frac{1}{2\lambda} C^2 -\frac{1}{\rho} G +\frac{1}{\xi} R^2
 \Bigr\}\,,
\end{split}
\end{equation}
where we have defined the square of the Weyl tensor $C^2$ and the four-dimensional Euler density $G$;
we have also neglected the boundary term $\Box R$.
The conformal limit of \eqref{eq:action4d} is obtained by taking $\xi\to\infty$.
This model was shown to be renormalizable a long time ago already \cite{Stelle:1976gc},
but, most importantly, it was shown that is \emph{can} be asymptotically free
\cite{Fradkin:1981iu,Avramidi:1985ki} if $\xi<0$.\footnote{%
Earlier computations included some mistakes,
we refer the reader to more recent papers, such as \cite{Salvio:2018crh,Jack:2020zgo}, for the correct results.}
The problem is that the same interaction is required to be positive, $\xi >0$,
by physically motivated constraints \cite{Salvio:2018crh} which are relevant also for cosmology \cite{Starobinsky:1980te}.
As it stands, for physical parameters, the model described by \eqref{eq:action4d} is thus
asymptotically free in $\lambda$, but Landau-trivial in $\xi$ (the integration
of the perturbative beta function would hit a Landau pole at a finite energy).

Nevertheless, the action \eqref{eq:action4d} has all the features that would make a natural complement to the Standard Model of particle physics, in fact
it compares well with the curvature square terms that give dynamic to the gauge fields, but also it is asymptotically free, at least in part.
For these reasons, the model was revived several times over the years \cite{deBerredoPeixoto:2004if,Salvio:2017qkx,Anselmi:2018ibi}.

Among its features, we notice that it has been proposed as an ultraviolet completion of quantum gravity and the standard model in the \emph{agravity} proposal \cite{Salvio:2017qkx}, which essentially suggests a regime beyond Planckian energies, $E$,
in which the renormalization group is controlled by \eqref{eq:action4d} for $E\gtrsim M_{\rm Pl}$ and its conformal version for $E \gg M_{\rm Pl}$. The decoupling of conformal models in the ultraviolet
is the mechanism that the agravity proposal has to avoid the perturbative Landau-pole.
However, the action \eqref{eq:action4d} has an additional problem in that it violates unitarity, unless some clever prescription generalizing the Wick rotation
is chosen to project away negative norm ghost modes \cite{Anselmi:2018ibi}, or a new symmetry is imposed on the spectrum \cite{Mannheim:2020ryw}.

The RG behavior of \eqref{eq:action4d} is known at the leading one loop order for the standard $\Diff$ realization,
but also for the $\Conf$ one in which there is no $R^2$ term \cite{deBerredoPeixoto:2003pj}. It would be interesting to extend these results to the next-to-leading order; furthermore, to the best of our knowledge, nobody has studied in depth the $\Diffstar$ realization \cite{Hamada:2002cm}, even though
the structure of the conformal anomaly in four dimensions is known to an extent similar to the previous case \cite{Antoniadis:1992xu}.
Since the couplings in \eqref{eq:action4d} all have Gaussian fixed points, it is customary to introduce the ratios $\omega=-3\lambda/\xi$ and $\theta=\lambda/\rho$, so that the perturbative series is controlled only by $\lambda$ (also seen through the expansion
$g_{\mu\nu}=\delta_{\mu\nu}+\sqrt{\lambda}h_{\mu\nu}$). For the standard diffeomorphisms symmetry, we have the beta functions
\begin{equation}\label{eq:beta4d}
\begin{split}
 &\beta_\lambda = -\frac{1}{(4\pi)^2}\frac{133}{10} \lambda^2
 \,,\qquad\quad
 \beta_\omega =-\frac{1}{(4\pi)^2}\frac{25+1098\omega+200\omega^2}{60}\lambda
 \qquad\quad
 \beta_\theta = \frac{1}{(4\pi)^2} \frac{7(56-171\theta)}{90}\lambda\,,
\end{split}
\end{equation}
instead for the conformal model ($\xi\to\infty$)
\begin{equation}\label{eq:beta4d-conf}
\begin{split}
 \beta_\lambda = -\frac{1}{(4\pi)^2}\frac{199}{15} \lambda^2
 \,,\qquad\quad
 \beta_\theta= \frac{1}{(4\pi)^2} \frac{261-796\theta}{60}\lambda\,.
\end{split}
\end{equation}
For $\lambda=0$ all betas are automatically zero, however the true fixed points are obtained by solving the system as $\lambda$ goes to zero, so in units of
a different RG ``time'', ${\rm d}\tilde{\mu}=\lambda {\rm d}\mu$.
A simple analysis shows that the first system has two nontrivial negative roots for $\omega$, one of which is clarly unphysical \cite{Groh:2011vn}, while the other one is argued to lead to an ill-defined Newtonian potential \cite{Salvio:2018crh}.
This means that the physically interesting action is supposed to have $\omega>0$,
where only $\lambda$ is asymptotically free, but $\xi$ is Landau-trivial \cite{Salvio:2017qkx}.
Nevertheless, the RG system induces nontrivial fractal properties for the effective structure of spacetime at small distances \cite{Becker:2019fhi}.

In relation to the discussion of the previous model in $d=2$, it would be tempting to ask whether the fixed points of the higher derivative model
have an interplay with the one which is ${\cal O}(\epsilon)$ in $d=2+\epsilon$. The answer is that we do not know, but evidence based on FRG methods suggest
that this is not the case, since both two- and four-dimensional fixed points can be seen in the same RG diagram
if quadratic divergences are taken into account \cite{Groh:2011vn}, though a more scheme independent statement would
be desirable.
Before concluding, it is interesting to notice that both $SU(N)$ gauge theories and general NLSMs
admit very similar higher derivative generalizations, see for example \cite{Casarin:2019aqw} and \cite{Percacci:2009fh}.
These higher derivative generalizations are also asymptotically free, at least in some couplings, and require a similar rescaling of the RG time by a
single coupling, say $\lambda$, which is thus required to control the perturbative series.

\subsection{Cubic gravity in $d=6$}\label{sect:6d}

We are approaching \emph{terra incognita}, because very little has been done in dimensions higher than four in relation
to the purposes of the guiding principles that we stated.
In six dimensions there are ten distinct operators that are constructed from three powers of the curvatures, modulo total derivatives.
We can parametrize the action as
\begin{equation}\label{eq:action6d}
\begin{split}
 S[g]=&
 \int {\rm d}^6 x \sqrt{g} \Bigl\{
 a_1 R \Box R + a_2 R_{\mu\nu} \Box R_{\mu\nu}
 + a_3 R^3
 + a_4 R R_{\mu\nu} R^{\mu\nu} 
 + a_5 R_{\mu}{}^{\nu} R_{\nu}{}^\alpha R_{\alpha}{}^{\mu}
  \\&
 + a_6 R_{\mu\nu}R_{\alpha\beta} C^{\mu\nu\alpha\beta}
 +a_7 R C_{\mu\nu\alpha\beta}C^{\mu\nu\alpha\beta}
 + a_8 R^{\mu\nu} C_{\mu\alpha\beta\gamma}C^{\nu\alpha\beta\gamma}
  \\&
 + a_9 C_{\mu\nu}{}^{\alpha\beta}C_{\alpha\beta}{}^{\rho\theta}C_{\rho\theta}{}^{\mu\nu}
 + a_{10} C^\mu{}_\alpha{}^\nu{}_\beta C^\alpha{}_\rho{}^\beta{}_\theta C^\rho{}_\mu{}^\theta{}_\nu
 \Bigr\}\,,
\end{split}
\end{equation}
and the couplings $a_i$ for $i=1,\cdots,10$ are correctly dimensionless in $d=6$. This action
is power counting renormalizable, so it is realistic to assume that the couplings $a_i$ are sufficient to remove the divergences
at least at one loop in the perturbative expansion, if not at all orders like in the previous case.

To state the obvious, the action \eqref{eq:action6d} evidently is much longer and more complicate than the two and four dimensional counterparts.
In fact, we believe that it has not been renormalized at one loop for the $\Diff$ or $\Diffstar$ realizations, yet.
Nevertheless, it has been considered in the $\Conf$ case using functional RG methods and background field approach in Ref.~\cite{Pang:2012rd}.
The conformally invariant action includes the manifestly invariant operators multiplying $a_9$ and $a_{10}$,
but also a linear combination of the others \cite{Pang:2012rd},
which can be related to a six dimensional $Q$-curvature.\footnote{
The $Q$-curvature needed here is often denoted $Q_{6,6}$, or $Q_{6,d}$ for the dimensional continuation \cite{Chernicoff:2018apt}.
In general $Q_{m,n}$ is the tensor defined as the constant part of an operator $P_{m,n}$ with leading part
$P_{m,n}= \Box^{\frac{m}{2}}+\cdots$
and that transforms as
$P_{m,n} \to {\rm e}^{-\frac{n+m}{2}\sigma} P_{m,n} {\rm e}^{\frac{n-m}{2}\sigma}$ for $g_{\mu\nu} \to {\rm e}^{2\sigma} g_{\mu\nu}  $.
Since the operator has ``nice'' transformation properties under Weyl rescalings, it is possible to construct an interesting
curvature from its constant part, $Q_{m,n}\equiv P_{m,n}1$, and couple it to topological charges.
In the examples of the previous sections, one can find relations of the monomials with $Q_{2,d}$ and $Q_{4,d}$ (modulo boundary terms),
which is a proof of concept for the utility of these curvatures in the context of critical models of quantum gravity.
}
However the Einstein's backgrounds for the metric that have been chosen in Ref.~\cite{Pang:2012rd} are not enough to uniquely assign a beta function to the
three couplings left, resulting in an incomplete RG flow, but the result is still impressive given the tensorial nature of the action.

An interesting aspect of \eqref{eq:action6d} is that it contains the notorious Goroff-Sagnotti term, $R_{\mu\nu}{}^{\alpha\beta}R_{\alpha\beta}{}^{\rho\theta}R_{\rho\theta}{}^{\mu\nu}$, which is used to argue the perturbative non-renormalizability
of the traditional Einstein-Hilbert action in dimension $d=4$ \cite{Goroff:1985th}.
The Goroff-Sagnotti term appears as a non-subtractable divergence when renormalizing \eqref{eq:action2d} at two loops in $d=4$,
and it was included in functional truncations for the first time in Ref.~\cite{Gies:2016con}.
Its tensor structure appears by expressing the Weyl tensor in terms of the Riemann tensor in the last two monomials. It is also
necessary to realize that in $d=4$ a totally antisymmetric expression with more than four indices is necessarily zero (one can have at most four antisymmetric indices through the tensor $\epsilon_{\mu\nu\rho\theta}$) so we can prove that
$R^\mu{}_\alpha{}^\nu{}_\beta R^\alpha{}_\rho{}^\beta{}_\theta R^\rho{}_\mu{}^\theta{}_\nu
=
\frac{1}{2}R_{\mu\nu}{}^{\alpha\beta}R_{\alpha\beta}{}^{\rho\theta}R_{\rho\theta}{}^{\mu\nu}$.

A complete RG analysis of \eqref{eq:action6d} would be an important theoretical achievement,
which could have strong implications as suggested by our discussion, through a potential interplay with the lower dimensional universality classes in their strongly interacting regimes, and which we hope is undertaken in the future.

\subsection{A final remark: unitarity}\label{sect:unitarity}

The discussion that we have presented in this section is based on the guiding principles that
we have listed in Sect.~\ref{sect:construction}. The core of our idea is to use the four guiding principles
constructively, but the resulting theories have to be tested for their physical validity.
The guiding principles themselves are based
on ideas coming from the theory of critical phenomena and are expressed in the language of the
renormalization group. One particular aspect of the theory of critical phenomena, especially
when applied to low dimensional condensed matter systems, is that it is not essential that
the considered models are unitary. In fact, a critical phenomenon can manifest in the mesoscopic description
of systems that otherwise have a physical built-in cutoff (for example, the scale of their microscopic components). Such systems do not have to be unitary, or, equivalently, reflection-positive
in case that they are Euclidean, because they are supposed to have a unitary completion
from their fundamental constituents.

On the other hand, ultraviolet complete theories of gravity, such as for example the conjectured asymptotically safe Einstein gravity, are not supposed to either have or need a further ultraviolet completion.
This is an especially delicate issue, because unitarity is not always easy to prove, both inside and outside
perturbation theory. We touched very briefly the problem of unitarity for the higher derivative model in Sect.~\ref{sect:4d}, but there is an extensive literature which recently has seen the influx of new ideas \cite{Anselmi:2018ibi,Mannheim:2020ryw,Arici:2017whq}. We do not explore this issue further, but
we dare add a fifth principle, upon which one should test the aforementioned theories.\footnote{%
Kindly suggested by a helpful referee.}

\bigskip

\noindent \fbox{
\parbox{0.972\textwidth}{
{\bf Guiding principle 5:} Any constructed critical theory that also wants to be an ultraviolet completion should be unitary.
}
}

\bigskip

\section{A toy-model: gauge theory in $d>4$}\label{sect:YM}

Now that we have listed the first few potential candidates for a universal theory of metric quantum gravity
following the reasonable guiding principles for their constructions at a given critical dimension, it is time to collect ideas on
what could be done with them.
Since its early inception, the idea of an asymptotically safe, ultraviolet complete, theory of quantum gravity in $d=4$
has been associated with the continuation to $\epsilon=2$ of the Einstein-Hilbert action in $d=2+\epsilon$ of Sect.~\ref{sect:2d}.
The most notable exceptions, in this sense, would be some early work of Niedermaier on the conjecture,
incorporating the perturbative RG of higher derivative gravity given in Sect.~\ref{sect:2d} as well as powerlaw
divergences to ``see'' a nontrivial RG for the lower derivative terms that are in a strong coupling regime \cite{Niedermaier:2009zz}, and the works of Falls discussing universal physical properties of asymptotic safety \cite{Falls:2015qga,Falls:2017cze,Falls:2018olk}.

Using the continuation of the theory in $d=2+\epsilon$ as a working hypothesis, we think that
the paramount problem is to assess its existence in $d=4$ and, more generally, in $d>2$.
The natural general question becomes: is it possible to extend an asymptotically free theory in a certain critical dimension $d_c$
and obtain an asymptotically safe theory for some natural-valued dimension $d>d_c$?
We have already given a known example, in the form of the NLSM, which is asymptotically free in $d_c=2$,
but also belongs to the same universality class as the linear model, that is nontrivial up to $d=4$.
In fact, the NLSM has been used as a toy-model for asymptotic safety in the past using functional methods \cite{Codello:2008qq,Percacci:2009fh},
however some conclusions seem unnatural; for example in Ref.~\cite{Codello:2008qq} it is predicted that
the NLSM has a nontrivial critical point for all $d>2$ in contrast with the above argument.
This is most likely caused by the lack of control of the operator expansion and cutoff parametric dependence,
which generally affects the application of functional methods to theories with nonlinearly realized symmetries
like the NLSM. The problem is shared, of course, by all metric theories of gravity that we listed.

As mentioned before, there are indications, based both on perturbation theory and Wilsonian methods
that there should be a conformal window for the existence of the critical point of $d=2+\epsilon$ gravity,
and that this window should extend above $d=4$, thus validating the conjecture \cite{Falls:2015qga,Gies:2015tca,Martini:2021slj, Labus:2015ska}.
In particular, Refs.~\cite{Falls:2015qga,Martini:2021slj} predict the conformal window $ d < 7.685$
from the leading order of perturbation theory, using two separate methods. The result might be subject to
radiative corrections, but it is further validated by qualitatively agreeing with
the functional approach with minimal dependence on the cutoff of Ref.~\cite{Gies:2015tca}.

It would be desirable, at this point, to have a toy-model, other than the NLSM, which exihibits similar properties
as the conjectured ones for gravity. The reason is that the NLSM is not entirely a perfect guiding principle, because we
know that its universality is captured by the one of the linear model for $d>2$.
The toy-model should also be under better control than $2d$ gravity, meaning that it should be known to a higher
extent as a quantum field theory, both perturbatively and nonperturbatively.
A simple candidate that fullfills all the requisites is a $SU(N)$ gauge theory: it is asymptotically free in $d=d_c=4$,
where its perturbative series is known to a reasonable amout of loops and where the growth of the coupling
towards the infrared makes the model approach a strongly-interacting phase. We could argue that
it should be asymptotically safe in $d=4+\epsilon$, at least for reasonably small values of $\epsilon$.

Earlier works have suggested, on the basis of the perturbative series, that the Yang-Mills gauge theory is
asymptotically safe above $d=4$, therefore it has a second order phase transition, but also that the conformal window
might extend up to $d\approx 6$ \cite{Peskin:1980ay,Morris:2004mg}. The critical properties change in such a way to
suggest that the transition might become first order at the effective upper critical dimension $d\approx 6$ \cite{Peskin:1980ay}.
To corroborate the end of the conformal window comes the picture suggested in Ref.~\cite{Gies:2003ic},
where it is shown, using functional RG methods, that, for small $\epsilon$, there is always an ultraviolet fixed point, but also
another nonperturbative fixed point besides it. For increasing values of $d$, and therefore of $\epsilon$,
the two fixed points collide leaving us with no zeroes for the beta function at $d \approx 6$.

To explain what is going on in Ref.~\cite{Gies:2003ic} from a different perspective,
imagine that in $d=4+\epsilon$ the square of the dimensionless gauge coupling has a beta function of the form
\begin{equation}\label{eq:beta-YM}
\begin{split}
 \mu \partial_\mu g^2 &= \epsilon g^2 + \beta_{g^2}\,,
\end{split}
\end{equation}
for some nonperturbative beta function $\beta_{g^2}$ in $d=d_c=4$.
The nonperturbative beta contains the leading order of perturbation theory, $\beta_{g^2}=-A g^4+\cdots$ and $A=\frac{22N}{3}$,
but also hides in the dots additional nonperturbative information coming from the strong coupling regime.
If we only use the perturbative part, there is always a solution $g^2\sim {\cal O}(\epsilon)$ of $\mu\partial_\mu g^2=0$.
However, if the nonperturbative part contributes so that $\beta_{g^2} \to +\infty $ for $g^2 \to +\infty$,
we must have at least another fixed point, and, for increasing $\epsilon$, this latter fixed point will collide with the perturbative one.\footnote{%
We are assuming here that there are no singularities in the beta function, so the form is assumed to be different from the one given
in \cite{Sannino:2010ue,Ryttov:2007cx}, for example.
}
The presence of a new fixed point is reminiscent of the Banks-Zaks one \cite{Banks:1981nn,Kaplan:2009kr}
and, as such, it leads to a more complicate phase-diagram, but with $d$ as ``parameter'' this time.
The collision would also explain why at $d=d_u\approx 6$ there should be the onset of a first order transition, in fact
the two fixed points should collide and form a complex conjugate pair, so we would expect that, slightly above
the effective upper critical dimension $d_u$, there is a weak first order transition (it is still affected by the complex pair of fixed point
that are close to the real axis).\footnote{More precisely, the transition should become infinite order exactly at $d=d_u$.} Sufficiently far away and above the effective critical dimension the transition would
become properly of first order.

\begin{figure}[tb]
\includegraphics[width=0.6\textwidth]{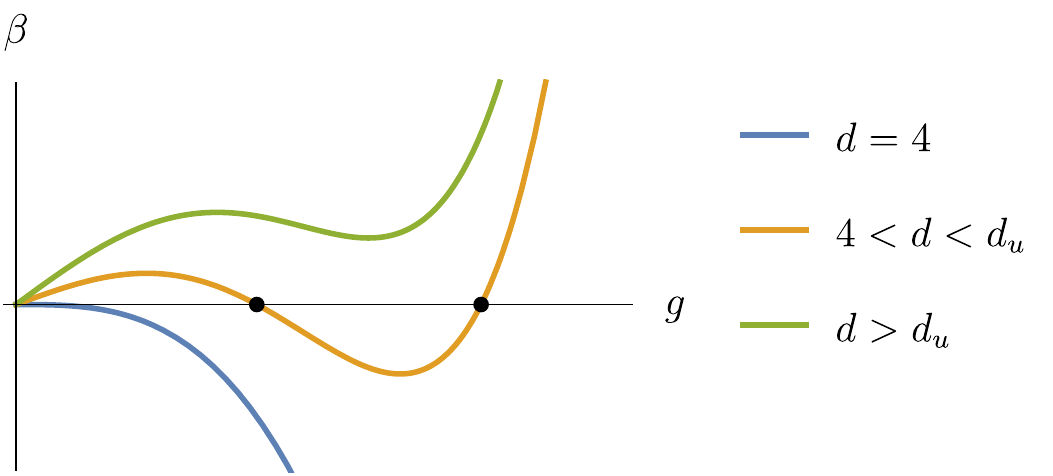}
\caption{
Conjectured behavior of the beta function of the Yang-Mills coupling in various dimensions $d$,
for which the slope at $g=0$ is proportional to $\epsilon=d-4$.
The blue function is the asymptotically free beta function of standard four dimensional Yang-Mills.
For an interval of dimensions between the critical value, $d=4$, and a speculative upper critical value, $d_u$,
the model admits two nontrivial fixed points, which are highlighted with dots in the figure. One is of ultraviolet nature,
while the other one is infrared and assumed to be associated to a multicritical theory in the main text.
Any change in $d$ corresponds to a change in the initial slope of the beta function,
so at $d=d_u$ the two fixed points collide and above $d_u$ the curve is not able to intercept the horizontal axis any longer \cite{Gies:2003ic}.
\label{fig:gaugebetas}}
\end{figure}

In order to validate or disprove the possibility that a $SU(N)$ gauge theory in $d>4$ is asymptotically safe,
the natural option would be to perform lattice simulations of some discretized action with $N=2$ and $d=5$,
which is the simplest configuration that is reasonably expected to have an ultraviolet fixed point and hence a second order critical point.
The use of the lattice Wilson's action with the traces of plaquettes in the fundamental representation
seems to lead only to a first order phase transition, but it could be argued that the inlusion of another bare parameter
could be enough to see that the first order point actually belongs to a first order line that terminates
in a second order point. The idea was explored early on in Ref.~\cite{Kawai:1992um}, but there was no clear signal of second order behavior,
though this could be related to the size of the lattices being a bit small for today's standards.
The problem has been readdressed recently in Ref.~\cite{Florio:2021uoz},
in which some signals of second order behavior are seen, though more work has to be done in this direction.

One interesting aspect of the search for the critical point in $d=5$ lattice $SU(2)$ Yang-Mills gauge theory
is that it shares striking similarities with the search for the lattice gravitational critical point performed both with
Euclidean and causal dynamical triangulations (EDT and CDT) \cite{Ambjorn:2000dv,Coumbe:2014nea}. In fact,
in both cases it is known that the Regge discretization with fixed triangles and total spacetime volume of the Einstein-Hilbert action
leads to a first order behavior if the only bare parameter is (related to) the Newton's constant. However,
if another bare parameter is added to the discretized action, a measure term in the case of EDT and a parameter measuring
microscopic causality in CDT, then there are signs of more complicate phase behaviors, which can include
a first order line (interpolating with the first order point of the case with only Newton's constant) that terminates
in a second order critical point. See also the comment given at the end of the caption of Fig.~\ref{fig:nlsmuniversality}
for further points on the relation between lattice simulations and quantum gravity.

These examples open up the possibility that it might simply be necessary to include one additional bare parameter
to see the critical point of either gravity or gauge theory above their respective critical dimensions.
It is tempting to speculate that the new operators that are introduced with the new bare parameter
are those responsible for the second order behavior and, in the RG description, must be related to the nature of the second nontrivial fixed point.
We know that the new operators must be in the strong coupling regime,
but they could still be related to a multicritical generalization of the Yang-Mills gauge theory that has a critical dimension different than four.
A candidate for such multicritical generalization has already appeared in the literature and has higher derivative interactions of the gauge field
with critical dimension $d_c=6$ \cite{Gracey:2015xmw,Casarin:2019aqw}.
This candidate can be found by following the same logic of our guiding principles,
but applying it to the gauge degrees of freedom, rather than to the metric.
An interesting aspect of this idea is that a higher multicriticality should emerge
from the competition of derivative interactions of various orders, which is known
to lead to critical field theories in the scalar case \cite{Safari:2017irw,Safari:2017tgs}.
Some of us plan to come back to this topic soon.

\section{Conclusions}\label{sect:conclusions}

We have tried to push-forward five reasonable guiding principles to attempt a classification
of critical theories which depend on a metric. The principles are inspired by ideas coming
from the application of field-theoretical methods in statistical mechanics
and are motivated by several examples. The resulting classification begins as perturbative,
in the sense that we classify critical theories on the basis of a critical dimension that is assumed to
allow for a perturbative definition of the field theory in its proximity, but becomes non-perturbative
when we argue that these critical theories could be extended away from their critical dimensions (and how).

Our main purpose is twofold. On the one hand, we want to suggest reasonable
candidate universality classes that can be relevant for the conjecture of asymptotic safety,
but also, on the other hand, we simply want to promote the role of more perturbative and universal arguments
in the literature of asymptotic safety.
In recent years, the discussion on the conjecture has been mainly based on the application
of functional RG methods and the Wetterich equation \cite{Wetterich:1992yh}, or, more appropriately,
its background version, which could be called the Reuter equation \cite{Reuter:1996cp}.

The main candidate for the critical model behind the conjecture is of course the extension
to $\epsilon=2$ of Einstein-Hilbert gravity in $d=2+\epsilon$,
which is known since the inception of the idea of asymptotic safety itself.
In an earlier work \cite{Martini:2021slj}, we discussed that the perturbative renormalization of this theory can be done close to two dimensions, but
all other instances of the dimensionality can be continued analytically, confirming
the presence of a conformal window in $d$ for the existence of the fixed point that can be computed perturbatively
and should include the physical case $d=4$ \cite{Falls:2015qga}.

However, we believe that, in such context, it is important to explain why there should be a conformal window.
In this contribution we have discussed a number of conjectures and similar open problems
that emerge from our analysis. In particular, we have discussed how the conformal window
could be caused by an interplay, happening in the strong-coupling regime,
between fixed points which have different critical dimensions (so they cannot be perturbative at the same time).
Our arguments in this direction are admittedly very conjectural,
but have been presented here in the hope of stimulating further research and discussions on these topics.

Even if very conjectural, our discussion has left out several important contributions
that are not covered by our classification, but still are important in the ecosystem of
the application of perturbative and field-theoretical methods to quantum gravity.
A prominent approach could be the one of Ref.~\cite{Modesto:2011kw} in which, following the results of \cite{Krasnikov:1987yj, Kuzmin:1989sp, Tomboulis:1997gg} a construction
of a nonlocal super-renormalizable theory in $d=4$ is attempted,
that of course differs from our suggested perturbatively (power-counting) renormalizable
models. Similar ideas have appeared in \cite{Biswas:2011ar} and \cite{Koshelev:2017tvv}.
We have not discussed changing the degrees of freedom either, for example introducing
a scalar field, which is known to be related to $f(R)$ gravity \cite{Ruf:2017bqx}.
Neither we explored alternative constructions of the path-integral,
such as the one ldeading to the unique effective action of Vilkovisky and DeWitt \cite{Vilkovisky:1984st}, which should
reduce drastically the gauge and parametric dependence of the final result and was recently applied to gravity \cite{Giacchini:2020zrl}.
The finite parts of the quantum effective action
of any theory of quantum gravity is well-known to include nonlocal form factors in the curvatures,
both in two \cite{Ribeiro:2018pyo} and four dimensions \cite{Franchino-Vinas:2018gzr, Knorr:2019atm, Codello:2015oqa},
which become very important in a cosmological setting \cite{Maggiore:2014sia,Codello:2015mba}.

\smallskip

\paragraph*{Acknowlegments.}
The research of AU was funded by the Deutsche Forschungsgemeinschaft under the Grant No.~Za 958/2-1.
OZ is grateful to C.~Bonati, A.~Codello, H.~Gies and R.~Percacci for inspiring discussions on these topics over the years.
We are especially grateful to H.~Gies for comments on an earlier version of the draft.
We are also grateful to a helpful referee suggesting the inclusion of unitarity as a guiding principle.
%



\end{document}